\documentstyle[sprocl,overcite,epsf]{article}
\def\Preprint{\vspace*{-5cm} %\noindent hep-ph/9701XXX
  \mbox{}       \hfill 
  FTUV/97-03 \\ \mbox{}\hfill 
  IFIC/97-03 \\  \mbox{}\hfill January 1997 \\ 
  \vspace{3.4cm}}
\def\refjl#1#2#3#4#5#6{\bibitem{#1} #2, {\it #3} {\bf #4} (#5) #6.}
\def\refbk#1#2#3#4{\bibitem{#1} #2, {\it #3}, #4.}
\def\etal{{\it et al}}
\def\NP{Nucl. Phys.}
\def\NPPS{Nucl. Phys. B (Proc. Suppl.)}
\def\PL{Phys. Lett.}
\def\PRL{Phys. Rev. Lett.}
\def\PR{Phys. Rev.}

\def\ZP{Z. Phys.}
\def\MPL{Mod. Phys. Lett.}

\newcommand{\eqn}[1]{(\ref{#1})}
\newcommand{\be}{\begin{equation}}
\newcommand{\ee}{\end{equation}}
\newcommand{\no}{\nonumber}
\newcommand{\bel}[1]{\be\label{#1}}
\newcommand{\ba}{\begin{array}{c}}
\newcommand{\bat}{\begin{array}{cc}}
\newcommand{\ea}{\end{array}}
\newcommand{\beqn}{\begin{eqnarray}}
\newcommand{\eeqn}{\end{eqnarray}}

\newcommand{\bi}{\begin{itemize}}
\newcommand{\ei}{\end{itemize}}

\newcommand{\rms}{\rm\scriptsize}

\newcommand{\lsim}{~{}_{\textstyle\sim}^{\textstyle <}~}

\newcommand{\cO}{{\cal O}}

\renewcommand{\thefootnote}{\fnsymbol{footnote}}

\begin{document}

\title{QCD TESTS FROM TAU DECAYS \footnote{
Invited talk at the 20$^{\rms th}$ Johns Hopkins Workshop:
{\it Non Perturbative Particle Theory} \& {\it Experimental Tests}
(Heidelberg, 27--29 June 1996)}}

\author{A. PICH}

\address{Departament de F\'{\i}sica Te\`orica, 
         IFIC,  CSIC --- Universitat de Val\`encia, \\ 
         Dr. Moliner 50, E--46100 Burjassot, Val\`encia, Spain}

\maketitle
\Preprint
\setcounter{footnote}{0}
\renewcommand{\thefootnote}{\alph{footnote}}

\abstracts{
The total $\tau$ hadronic width can be accurately calculated using analyticity
and the operator product expansion. 
The theoretical analysis of this observable is updated to
include all available perturbative and non-perturbative corrections. 
The experimental determination
of $\alpha_s(M_\tau^2)$ and its actual uncertainties are discussed.}

\section{Introduction}

The inclusive character of the total $\tau$ hadronic width renders possible 
an accurate calculation of the ratio
\cite{BR:88,NP:88,ORSAY:90,BNP:92,LDP:92a,OHIO:92,QCD:94,NA:95,BR:96}
[$(\gamma)$ represents additional photons or lepton pairs]
\be\label{eq:r_tau_def}
     R_\tau \equiv { \Gamma [\tau^- \rightarrow \nu_\tau
                   \,\mbox{\rm hadrons}\, (\gamma)] \over
                         \Gamma [\tau^- \rightarrow
                \nu_\tau e^- {\bar \nu}_e (\gamma)] } ,
\ee
using standard field theoretic methods. 
If strong and electroweak radiative corrections are
ignored and if the masses of final--state particles are neglected,
the universality of the W
coupling to the fermionic charged currents implies 
\be\label{eq:naive} R_\tau \, \simeq \, N_c \,
 (|V_{ud}|^2 + |V_{us}|^2) \, \simeq \, 3 \; ,  
\ee
which compares quite well with the experimental average \cite{tau96}
$R_\tau =3.649 \pm 0.014$.
This provides strong evidence for the colour degree of freedom $N_c$.

The QCD dynamics is able to account quantitatively for the 
20\% difference 
between the na\"{\i}ve prediction (\ref{eq:naive})
and the measured value of $R_\tau$. Moreover,
the uncertainties in the theoretical calculation of $R_\tau$ are
quite small. The value of $R_\tau$ can then be accurately predicted
as a function of $\alpha_s(M_\tau^2)$.
Alternatively,  measurements of inclusive $\tau$ decay rates
can be used to determine the value of the QCD running coupling
$\alpha_s(M_\tau^2)$ at the scale of the $\tau$ mass.
In fact, $\tau$ decay is probably the lowest--energy process from which the
running coupling constant can be extracted cleanly, without hopeless
complications from non-perturbative effects.  The $\tau$ mass \cite{tau96}, 
$M_\tau = (1777.00{\,}^{+0.30}_{-0.27})$  MeV, 
lies fortuitously
in a compromise region  where the coupling constant
$\alpha_s(M_\tau^2)$ is large enough that $R_\tau$ is sensitive to its
value, yet still small enough that the perturbative expansion
still converges well. Moreover, 
the non-perturbative contributions to the total $\tau$ hadronic width are
very small.
 
It is the inclusive nature of the total semihadronic decay rate that makes
a rigorous theoretical calculation of $R_\tau$ possible.
The only separate contributions to $R_\tau$ that can be calculated are those
associated with specific quark currents.  We can calculate the
non-strange and strange contributions to $R_\tau$, and resolve these
further into  vector and axial--vector contributions.
Since strange decays cannot be resolved experimentally into vector and 
axial--vector  contributions,
we will decompose our predictions for $R_\tau$
into only three categories:
\be\label{eq:r_tau_v,a,s}
 R_\tau \, = \, R_{\tau,V} + R_{\tau,A} + R_{\tau,S}\, .
\ee
Non-strange semihadronic decays of the $\tau$ are resolved experimentally
into vector ($R_{\tau,V}$) and axial--vector ($R_{\tau,A}$)
contributions according to whether the
hadronic final state includes an even or odd number of pions.
Strange decays ($R_{\tau,S}$) are of course identified by the
presence of an odd number of kaons in the final state.
The na\"{\i}ve predictions for these three ratios are
$R_{\tau,V} \simeq R_{\tau,A} \simeq (N_c/2)|V_{ud}|^2$ and
$R_{\tau,S} \simeq N_c |V_{us}|^2$, which add up to (\ref{eq:naive}).

\section{Theoretical Framework}  %{THEORETICAL FRAMEWORK}

The theoretical analysis of $R_\tau$ involves
the two--point correlation functions for
the vector $\, V^{\mu}_{ij} = \bar{\psi}_j \gamma^{\mu} \psi_i \, $
and axial--vector
$\, A^{\mu}_{ij} = \bar{\psi}_j \gamma^{\mu} \gamma_5 \psi_i \,$
colour--singlet quark currents ($i,j=u,d,s$):
\beqn\label{eq:pi_v}
\Pi^{\mu \nu}_{ij,V}(q) &\!\!\! \equiv &\!\!\!
 i \int d^4x \, e^{iqx} 
\langle 0|T(V^{\mu}_{ij}(x) V^{\nu}_{ij}(0)^\dagger)|0\rangle  ,
\\ 
\label{eq:pi_a}
\Pi^{\mu \nu}_{ij,A}(q) &\!\!\! \equiv &\!\!\!
 i \int d^4x \, e^{iqx} 
\langle 0|T(A^{\mu}_{ij}(x) A^{\nu}_{ij}(0)^\dagger)|0\rangle  .
\eeqn
%
%The vector ($V$) and axial--vector ($A$) correlators 
%These correlators
They have the Lorentz decompositions
\bel{eq:lorentz}
\Pi^{\mu \nu}_{ij,V/A}(q)  = 
  (-g^{\mu\nu} q^2 + q^{\mu} q^{\nu}) \, \Pi_{ij,V/A}^{(1)}(q^2) 
  +   q^{\mu} q^{\nu} \, \Pi_{ij,V/A}^{(0)}(q^2) ,
\ee
where the superscript $(J)$
denotes the angular momentum 
$J=1$ or $J=0$ in the hadronic rest frame.

%%%%%%%%%%%%%%%%%% FIGURE %%%%%%%%%%%%%%%%%%%%%
\begin{figure}[tbh]
\label{fig:circle}
\centerline{\epsfxsize =7.2cm \epsfbox{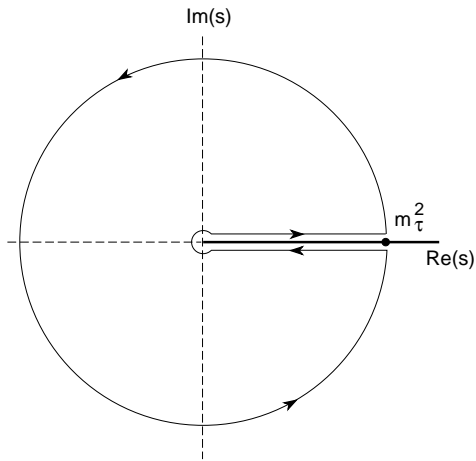}}  %8cm
\vspace{-0.5cm}
\caption{Integration contour in the complex $s$ plane, used to obtain
Eq.~\protect\eqn{eq:circle}}
\end{figure}
%%%%%%%%%%%%%% END FIGURE %%%%%%%%%%%%%%%%%%%%%

The imaginary parts of the two--point functions 
$\, \Pi^{(J)}_{ij,V/A}(q^2) \, $ 
are proportional to the spectral functions for hadrons with the corresponding
quantum numbers.  The semihadronic decay rate of the $\tau$
can be written as an integral of these spectral functions
over the invariant mass $s$ of the final--state hadrons:
\bel{eq:spectral}
R_\tau   =  
12 \pi \int^{M_\tau^2}_0 {ds \over M_\tau^2 } \,
 \left(1-{s \over M_\tau^2}\right)^2 
\biggl[ \left(1 + 2 {s \over M_\tau^2}\right) 
 \mbox{\rm Im} \Pi^{(1)}(s)
 + \mbox{\rm Im} \Pi^{(0)}(s) \biggr]  . 
\ee
The appropriate combinations of correlators are 
\bel{eq:pi}
\Pi^{(J)}(s)  \equiv
  |V_{ud}|^2 \, \left( \Pi^{(J)}_{ud,V}(s) + \Pi^{(J)}_{ud,A}(s) \right)
+ |V_{us}|^2 \, \left( \Pi^{(J)}_{us,V}(s) + \Pi^{(J)}_{us,A}(s) \right). 
\ee
The contributions coming from the first two terms correspond to 
$R_{\tau,V}$ and $R_{\tau,A}$ respectively, while  
$R_{\tau,S}$ contains the remaining Cabibbo--suppressed contributions.

Since the hadronic spectral functions are sensitive to the non-perturbative
effects of QCD that bind quarks into hadrons, the integrand in 
Eq.~(\ref{eq:spectral}) cannot be calculated at present from QCD.
Nevertheless the integral itself can be calculated systematically
by exploiting
the analytic properties of the correlators $\Pi^{(J)}(s)$.
They are analytic
functions of $s$ except along the positive real $s$ axis, where their
imaginary parts have discontinuities.  The integral (\ref{eq:spectral}) can
therefore be expressed as a contour integral 
in the complex $s$ plane running
counter--clockwise around the circle $|s|=M_\tau^2$:
\bel{eq:circle}
R_\tau  =
6 \pi i \oint_{|s|=M_\tau^2} {ds \over M_\tau^2} \,
 \left(1 - {s \over M_\tau^2}\right)^2
 \left[ \left(1 + 2 {s \over M_\tau^2}\right) \Pi^{(0+1)}(s)
         - 2 {s \over M_\tau^2} \Pi^{(0)}(s) \right] . 
\ee

The advantage of expression (\ref{eq:circle})  
over (\ref{eq:spectral})   %for $R_\tau$
is that it requires the correlators only for
complex $s$ of order $M_\tau^2$, which is significantly larger than the scale
associated with non-perturbative effects in QCD.  The short--distance
Operator Product Expansion (OPE) can therefore be used to organize
the perturbative and non-perturbative contributions
to the correlators into a systematic expansion \cite{SVZ:79}
in powers of $1/s$,
\be\label{eq:ope}
 \Pi^{(J)}(s) = \sum_{D=2n}\,\sum_{\mbox{\rms dim} {\cal O} = D}
 {{\cal C}^{(J)}(s,\mu) \,\langle {\cal O}(\mu)\rangle\over (-s)^{D/2}} ,
\ee
 where the inner sum is over local gauge--invariant
 scalar operators of dimension $D=0,2,4,\ldots $
 The possible uncertainties associated with the use of the OPE near the
 time--like axis are absent in this case, because
 the integrand in Eq. (\ref{eq:circle}) includes a factor
 $(1- s/M_\tau^2)^2$, which provides a double zero at $s=M_\tau^2$,
 effectively suppressing the contribution from the
 region near the branch cut.
 The parameter $\mu$ in Eq. (\ref{eq:ope})
 is an arbitrary factorization scale, which separates long--distance
 non-perturbative effects, which are absorbed into the vacuum matrix elements
 $\langle {\cal O}(\mu)\rangle $, from short--distance effects, which belong
 in the Wilson coefficients ${\cal C}^{(J)}(s,\mu)$.
 The $D=0$ term (unit operator) corresponds to the pure perturbative
contributions, neglecting quark masses. The leading quark--mass
corrections     generate the $D=2$ term. The first dynamical operators
involving non-perturbative physics appear at $D=4$.
Inserting the functions (\ref{eq:ope})
 into (\ref{eq:circle}) and evaluating the contour integral, $R_\tau$
 can be expressed as an expansion in powers of $1/M_\tau^2$,
 with coefficients that depend only logarithmically on $M_\tau$.

It is convenient to express the corrections to $R_\tau$
from dimension--$D$ operators in terms of the
fractional corrections $\delta^{(D)}_{ij,V/A}$ to the 
na\"{\i}ve contribution
from the current with quantum numbers $ij,V$ or $ij,A$:
\beqn\label{eq:r_v}
R_{\tau,V/A} &\!\!\! = &\!\!\! {3 \over 2} |V_{ud}|^2 
   S_{EW} \left( 1 + \delta_{EW}' + 
      \sum_{D=2n} \delta^{(D)}_{ud,V/A} \right) , 
\no \\ \label{eq:r_s} 
R_{\tau,S} &\!\!\! = &\!\!\!
 3 \, |V_{us}|^2 S_{EW} \left( 1 + \delta_{EW}' +
  \sum_{D=2n} \delta^{(D)}_{us} \right)  .                                               
\eeqn
$\delta^{(D)}_{ij} = (\delta^{(D)}_{ij,V} + \delta^{(D)}_{ij,A})/2$
is the average of the vector and axial--vector corrections.
The dimension--0 contribution is the
purely perturbative correction neglecting quark masses,
which is the same for all the components of $R_\tau$:
$\delta^{(0)}_{ij,V/A} = \delta^{(0)}$.
The factors \cite{MS:88}
\bel{eq:s_ew}
 S_{EW}  = 
\left( {\alpha(m_b^2) \over \alpha(M_\tau^2)} \right)^{9\over19}
         \left( {\alpha(M_W^2) \over \alpha(m_b^2)} \right)^{9\over 20}
         \left( {\alpha(M_Z^2) \over \alpha(M_W^2)} \right)^{36\over 17}
      =  1.0194  
\ee
and  \cite{BL:90}
\bel{eq:d_ew'}
 \delta_{EW}' = {5\over 12} \, {\alpha(M_\tau^2)\over\pi } \simeq 0.0010 \, ,
\ee
contain the known electroweak corrections.

Adding the three terms, the total ratio $R_\tau$ is
\beqn\label{eq:r_total}
\lefteqn{
R_{\tau}  =  3 \left( |V_{ud}|^2 + |V_{us}|^2 \right)
S_{EW} \biggl\{ 1 + \delta_{EW}' + \delta^{(0)}}
\no\\ && \qquad\qquad\qquad\qquad\mbox{}
 + \sum_{D=2,4,...}
         \left( \cos^2 \theta_C \delta^{(D)}_{ud}
         + \sin^2 \theta_C \delta^{(D)}_{us} \right) \biggr\} ,
\eeqn
where 
$\sin^2\theta_C\equiv |V_{us}|^2/(|V_{ud}|^2 + |V_{us}|^2)$.

\section{Perturbative Corrections}
%\section{PERTURBATIVE CORRECTIONS}

In the chiral limit ($m_u=m_d=m_s=0$),
the vector and axial--vector currents are conserved.
This implies  $s \Pi^{(0)}(s) = 0$; therefore, only the correlator 
$\Pi^{(0+1)}(s)$
 contributes to Eq.~(\ref{eq:circle}). 
Owing \cite{TR:79,AN:79} to the chiral invariance of massless QCD, 
$\Pi^{(0+1)}_{ij,V}(s) = \Pi^{(0+1)}_{ij,A}(s) \equiv \Pi(s)$
($i\not=j$) at any finite order in $\alpha_s$. 

The result is more conveniently expressed in terms of the 
logarithmic derivative of the two--point correlation 
function of the vector (axial) current, 
\be\label{eq:d}
D(s)  \equiv  - s {d \over ds } \Pi(s)  
=  {1\over 4 \pi^2} \sum_{n=0}  K_n
\left( {\alpha_s(s)\over \pi}\right)^n  , 
\ee
which satisfies an homogeneous renormalization--group equation.
The $K_n$ coefficients are known \cite{ChKT:79,DS:79,CG:80,GKL:91,SS:91}
to order $\alpha_s^3$.
For $n_f=3$ quark flavours, one has:
$$
K_0  =  K_1 = 1\, , 
$$
\bel{eq:Kfactors}
K_2  =  {299\over 24} - 9 \zeta(3) = 1.63982\, , 
\ee
$$
K_3(\overline{MS})  = 
{58057\over 288} - {779\over 4}\zeta(3) + {75\over 2} \zeta(5) 
= 6.37101 \, .
$$

The perturbative component of $R_\tau$ is given by
\be\label{eq:r_k_exp}
{R_\tau^{\mbox{\rms pert}}\over (|V_{ud}|^2 + |V_{us}|^2) S_{EW}}
\equiv 3 \,\{ 1 + \delta^{(0)} \}  =  3
\sum_{n=0}  K_n \, A^{(n)}(\alpha_s) \, ,
\ee
where
the functions \cite{LDP:92a}
\bel{eq:a_xi}
A^{(n)}(\alpha_s)  =  {1\over 2 \pi i}
\oint_{|s| = M_\tau^2} {ds \over s} \,
  \left({\alpha_s(-s)\over\pi}\right)^n  
 \left( 1 - 2 {s \over M_\tau^2} + 2 {s^3 \over M_\tau^6}
         - {s^4 \over  M_\tau^8} \right)  
\ee
are contour integrals in the complex plane,
which only depend on $\alpha_s(M_\tau^2)$.

The running coupling $\alpha_s(-s)$ in Eq. (\ref{eq:a_xi})
can be expanded in powers of
$\alpha_s(M_\tau^2)$, with coefficients that are polynomials in
$\log{(-s/M_\tau^2)}$.
The perturbative expansion of $\delta^{(0)}$ in powers of
$a_\tau\equiv\alpha_s(M_\tau^2)/\pi$  then takes
the form \cite{BR:88,NP:88,ORSAY:90,BNP:92,LDP:92a,OHIO:92,QCD:94}:
\beqn\label{eq:delta_0}
\delta^{(0)} &\!\!\! = &\!\!\!
  \sum_{n=1} (K_n + g_n) \, a_\tau^n,
        %\left(\alpha_s(M_\tau^2)/\pi\right)^n
\no\\ &\!\!\! = &\!\!\!
  a_\tau %{\alpha_s(M_\tau^2)\over\pi}
  + \left( K_2 - {19\over 24} \beta_1 \right) 
  a_\tau^2  %\left( {\alpha_s(M_\tau^2)\over\pi}\right)^2
\\ &\!\!\! &\!\!\!
  + \left( K_3 - {19 \over 12} K_2 \beta_1
  - {19 \over 24} \beta_2
         + {265 - 24 \pi^2\over 288} \beta_1^2 \right) 
  a_\tau^3  %\left( {\alpha_s(M_\tau^2)\over\pi}\right)^3
  + \, \cO (a_\tau^4)  
\no\\ &\!\!\! = &\!\!\!
  a_\tau   %  {\alpha_s(M_\tau^2)\over\pi}
  + 5.2023 \, a_\tau^2  % \left( {\alpha_s(M_\tau^2)\over\pi}\right)^2
  + 26.366 \, a_\tau^3  %\left( {\alpha_s(M_\tau^2)\over\pi}\right)^3
       + \, \cO(a_\tau^4)  \, .\no
\eeqn        

The complex integration along the circle $|s|=M_\tau^2$
generates the $g_n$ coefficients,
which depend on $K_{m<n}$
and on $\beta_{m<n}$, where
\bel{eq:beta}
\beta_1=-9/2 \, , \qquad
\beta_2=-8 \, , \qquad
\beta_3(\overline{MS})=-3863/192 \, , \qquad\ldots
\ee
%
%[$\beta_1=-9/2$, $\beta_2=-8$, $\beta_3(\overline{MS})=-3863/192$, ..., 
are the coefficients of the QCD $\beta$ function for $n_f=3$.
One observes  \cite{LDP:92a} that the
$g_n$ contributions are larger than the original $K_n$
coefficients ($g_2  = 3.5625$, $g_3 = 19.9949$).
For instance, the bold--guess value
$K_4 \sim K_3 (K_3/K_2)\approx 25$
is to be compared with $g_4=78.0029$.
These large corrections give rise to a sizeable
renormalization scale dependence \cite{LDP:92a,CHKL:91}
of the truncated $\cO(a_\tau^3)$ result. 
The reason of such uncomfortably large  contributions \cite{LDP:92a} stems  from
the long running along the circle $s=M_\tau^2\exp{(i\phi)}$
($\phi\epsilon [0, 2\pi]$) in Eq. (\ref{eq:a_xi}).
When the running coupling $\alpha_s(-s)$
is expanded in powers of $\alpha_s(M_\tau^2)$, one
gets imaginary logarithms, $\log{(-s/M_\tau^2)} = i (\phi - \pi)$,
which are large in some parts of the integration range.
The radius of convergence of this expansion is actually quite small.
A numerical analysis of the series \cite{LDP:92a}
shows that, at the three--loop level,
an upper estimate for the convergence radius  is
$a_{\tau,\mbox{\rms conv}} \, < 0.11$.

Note, however,
that there is no deep reason to stop the  $A^{(n)}(\alpha_s)$ integral
expansions at ${\cal O}(\alpha_s^3)$.
One can calculate
the $A^{(n)}(\alpha_s)$ expansion  to all orders in 
$\alpha_s$, apart from the unknown
$\beta_{n>3}$ contributions, which are
likely to be negligible.
Even for $a_\tau$ larger
than the radius of convergence $a_{\tau,\mbox{\rms conv}}$, the integrals
$A^{(n)}(\alpha_s)$ are well--defined functions that can be numerically computed,
 by using in Eq. (\ref{eq:a_xi}) the exact
solution for $\alpha_s(-s)$ obtained from the renormalization--group 
$\beta$--function equation.
Thus a more appropriate approach \cite{LDP:92a,PIV:92} is to
use a $K_n$ expansion of $R_\tau^{\mbox{\rms pert}}$ as in
Eq.~(\ref{eq:r_k_exp}),
and to fully keep the known three--loop calculation of the functions
$A^{(n)}(\alpha_s)$. The perturbative uncertainties are then reduced to the
corrections coming from the unknown $\beta_{n>3}$ and $K_{n>3}$
contributions, since the $g_n$ contributions are properly resummed
to all orders.
To appreciate the size of the effect, Table~\ref{tab:a} 
gives the exact results \cite{LDP:92a}
for $A^{(n)}(\alpha_s)$ ($n=1,2,3$) obtained at the one--, two-- and three--loop
approximations (i.e. $\beta_{n>1}=0$, $\beta_{n>2}=0$,
 and $\beta_{n>3}=0$,
respectively), together with the final value of
$\delta^{(0)}$, 
for
$a_\tau=0.1$. For comparison, the numbers coming from the truncated
expressions at order $a_\tau^3$ are also given.
Although the difference between the
exact and truncated results represents a tiny
$0.7\% $ effect on $R_\tau$,
it produces a sizeable $4\% $ shift on the value of $\delta^{(0)}$.
The $\delta^{(0)}$ shift, which reflects into a corresponding
shift in the experimental $\alpha_s(M_\tau^2)$ determination,
depends strongly on the value of the coupling constant; for
$a_\tau=0.14$ the $\delta^{(0)}$ shift reaches the $-20\%$ level.

%%%%%%%%%%%%%%%%%%%%%%%% TABLE 1 %%%%%%%%%%%%%%%%%%%%%%%%%%%%%%%
%
\begin{table}[tbh]
\caption{
Exact results
for $A^{(n)}(\alpha_s)$ ($n=1,2,3$) obtained at the one-- ($\beta_{n>1}=0$),
two-- ($\beta_{n>2}=0$) and three--loop ($\beta_{n>3}=0$)
approximations, together with the final value of
$\delta^{(0)}$, for
$a_\tau=0.1$. For comparison, the numbers coming from the truncated
expressions at order $a_\tau^3$ are also given. \hfill\mbox{} }
\label{tab:a}
\vspace{0.4cm}\centering
\begin{tabular}{|c|c|c|c|c|}
\hline  
Loops & $A^{(1)}$ & 
$A^{(2)}$ & $A^{(3)}$ &
$\delta^{(0)}$ 
\\ \hline  
$1$ & $0.13247$ & $0.01570$ & $0.00170$ & $0.1690$ \\
$2$ & $0.13523$ & $0.01575$ & $0.00163$ & $0.1714$ \\
$3$ & $0.13540$ & $0.01565$ & $0.00160$ & $0.1712$ \\
\hline   
$\cO(\alpha_s^3)$ &
   $0.14394$ & $0.01713$ & $0.00100$ & $0.1784$ 
\\ \hline
\end{tabular}
\end{table}
%
%%%%%%%%%%%%%%%%%%%%%%%% END TABLE %%%%%%%%%%%%%%%%%%%%%%%%%%%%%%

Notice that the difference between using the one-- or two--loop approximation to
the $\beta$ function is already quite small
($1.4\%$ effect on  $\delta^{(0)}$), while the change induced by the
three--loop corrections is completely negligible ($0.1\%$). Therefore (unless
the $\beta$ function has some unexpected pathological behaviour at higher
orders), the error induced by the truncation of the $\beta$ function at third
order should be smaller than $0.1\% $ and therefore can be safely neglected.
The sensitivity on the choice of renormalization scale and
renormalization scheme is also very small \cite{LDP:92a,RS:96}.

%%%%%%%%%%%%%%%%%%%%%%%% TABLE 2 %%%%%%%%%%%%%%%%%%%%%%%%%%%%%%%
%
\begin{table}[thb]
\centering
\caption{$\delta^{(0)}$ for different values of $\alpha_s(M_\tau^2)$.}
\label{tab:perturbative}
\vspace{0.4cm}
\begin{tabular}{|c|c|c|c|}            %{cc|cc}  %||c|c||c|c|}
\hline  
$\alpha_s(M_\tau^2)$ & \multicolumn{2}{c|}{$\delta^{(0)}$} &
$\Delta(\delta^{(0)})$ 
\\ \cline{2-3}
& $K_4 = 0$ & $K_4=27.5$ &
\\ \hline
$0.30$ & $0.161$ & $0.164$ & $\pm0.006$
\\
$0.31$ & $0.168$ & $0.172$ & $\pm0.007$
\\
$0.32$ & $0.176$ & $0.180$ & $\pm0.008$
\\
$0.33$ & $0.183$ & $0.188$ & $\pm0.008$
\\
$0.34$ & $0.191$ & $0.196$ & $\pm0.009$
\\
$0.35$ & $0.198$ & $0.203$ & $\pm0.010$
\\
$0.36$ & $0.205$ & $0.211$ & $\pm0.010$
\\       
$0.37$ & $0.213$ & $0.219$ & $\pm0.011$
\\
$0.38$ & $0.220$ & $0.226$ & $\pm0.012$
\\
$0.39$ & $0.227$ & $0.234$ & $\pm0.012$
\\
$0.40$ & $0.234$ & $0.241$ & $\pm0.013$
\\  \hline   
\end{tabular}
\end{table}
%
%%%%%%%%%%%%%%%%%%%%%%%% END TABLE %%%%%%%%%%%%%%%%%%%%%%%%%%%%%%

The dominant perturbative uncertainties come  from
the unknown higher--order coefficients $K_{n>3}$.
The $\cO(\alpha_s^4)$ contribution has been estimated \cite{KS:95}
using scheme--invariant methods, namely 
the principle of minimal sensitivity \cite{ST:81}
and the effective charge approach \cite{GR:80}, with the
result \cite{KS:95}:
\bel{eq:k4}
K_4^{\mbox{\rms est}} = 27.5 \, .
\ee
This number is very close to the na\"{\i}ve guess \cite{LDP:92a}
$K_4 \sim (K_3/K_2) K_3 \approx 25$.
A similar estimate,
$K_4^{\mbox{\rms NNA}} = 24.8$, 
is obtained 
\cite{BR:93,BRK:93,BE:93,LTM:94,BB:95,BBB:95,NE:96}
in the limit of a large number of quark
flavours, using the so--called {\it naive non-abelianization}
prescription \cite{BG:95}
($n_f \to 3\beta_1 = n_f -{33\over 2} = -{27\over 2}$).
From a fit to the experimental $\tau$ data, the value
$K_4^{\mbox{\rms fit}}=29\pm 5$ has been also quoted \cite{LD:94}.

Using the estimate \eqn{eq:k4}, the $\cO(\alpha_s^4)$ correction
amounts to a 0.004 increase of $\delta^{(0)}$ for $a_\tau=0.1$.
The resulting perturbative contribution $\delta^{(0)}$, obtained through
Eqs. (\ref{eq:r_k_exp}) and (\ref{eq:a_xi}), is given in
Table~\ref{tab:perturbative} for different values of 
the strong coupling constant $\alpha_s(M_\tau^2)$.
In order to be conservative, and to account for all possible sources
of perturbative uncertainties,
we have  used \cite{OHIO:92,QCD:94}
\bel{eq:perror} 
\Delta(\delta^{(0)}) = \pm 50 \, A^{(4)}(\alpha_s)\, ,
\ee
as an estimate of the theoretical error on $\delta^{(0)}$.
Note that, for the relevant values of $\alpha_s$, this is
of the same size as $K_3\, A^{(3)}(\alpha_s)$; thus, this
error estimate is conservative enough to apply \cite{QCD:94}
in the worst possible scenario,
where the onset of the asymptotic behaviour of the
perturbative series were already reached for $n=3,4$.

There have been attempts
\cite{BE:93,LTM:94,BB:95,BBB:95,NE:96}
to improve the perturbative prediction
by performing an all--order summation a certain class
of higher--order corrections (the so-called ultraviolet
renormalon chains). This can be accomplished using
exact large--$n_f$ results and applying the
{\it naive non-abelianization} prescription \cite{BG:95}.
Unfortunately, the naive resummation turns out to be renormalization--scheme
dependent beyond one loop \cite{LTM:94,CH:95}.
More recently, a renormalization--scheme--invariant summation has been
presented \cite{MT:96}.
The final effect of the higher--order corrections (beyond $K_4$)
turns out to be small.\footnote{
The present results do not include yet the resummation of the $g_n$
running corrections through the $A^{(n)}(\alpha_s)$ functions.}

\section{Power Corrections}

The $1/M_\tau^2$ contributions $\delta^{(2)}_{ij}$ to the ratio $R_\tau$
are simply the leading quark--mass corrections to the perturbative
QCD result of the previous section. 
These contributions are known \cite{BNP:92,CK:93}
to order $\alpha_s^2$.
Quark--mass corrections are certainly tiny for the up and down quarks
($\delta^{(2)}_{ud}\sim -0.08\% $), but 
the correction from the strange--quark mass is important
for strange decays \cite{BNP:92,CK:93}: ($m_u=m_d=0$)
\bel{eq:delta_dos}
\delta^{(2)}_{us,V/A} = -8 {\overline m_s^2\over M_\tau^2}
\left\{1 + {16 \over 3} a_\tau  + 11.03\, a_\tau^2\right\} ,
\ee 
where $\overline m_s\equiv m_s(M_\tau^2)$ 
is the running mass of the strange quark
evaluated at the scale $M_\tau$.   
For $\alpha_s(M_\tau^2) = 0.35$, $\delta^{(2)}_{us}\approx  -20\% $;
nevertheless, because of the $\sin^2{\theta_C}$ suppression, the
effect on the total ratio $R_{\tau}$  is only $-(1.0\pm 0.2) \%$.

Since the $\tau$ mass is a quite low energy scale, we should
worry about possible non-perturbative effects.  %contributions.
In the framework of the OPE, the long--distance dynamics is absorbed
into the vacuum matrix elements 
$\langle {\cal O}(\mu)\rangle $, which are (at present) quantities
to be fixed phenomenologically.
If the logarithmic dependence of the Wilson coefficients
${\cal C}^{(J)}(s,\mu)$
on $s$ is neglected
(this is an effect of order $\alpha_s^2$),
the contour integrals 
can be evaluated trivially using Cauchy's residue theorem,
and are non-zero only for $D= 2, 4, 6$ and $8$.
The corrections simplify even further if we also
take the chiral limit ($m_u =m_d =m_s =0$).
The dimension--2 corrections then vanish because
there are no operators of dimension 2.  In the chiral limit,
$s \Pi^{(0)}(s) =0$;
thus only the $\Pi^{(0+1)}(s)$ term in 
Eq.~(\ref{eq:circle}) contributes to $R_\tau$.
The form of  the kinematical factor multiplying $\Pi^{(0+1)}(s)$
in Eq.~(\ref{eq:circle})  is such that, when the $s$--dependence of the
Wilson coefficients is ignored,
only the $D=6$ and $D=8$ contributions survive the integration.
The power corrections to $R_\tau$  then reduce to 
\cite{BR:88,NP:88,ORSAY:90,BNP:92}
\beqn\label{eq:np_naive}
  \delta^{(6)}_{ij,V/A} &\!\!\! \simeq &\!\!\!  - 24 \pi^2
      { \left[ \sum {\cal C}^{(0+1)}_{ij,V/A} 
\langle{\cal O}\rangle \right]^{(D=6)}
         \over M_\tau^6}\, , 
\no\\
  \delta^{(8)}_{ij,V/A} &\!\!\! \simeq &\!\!\! - 16 \pi^2
     { \left[ \sum {\cal C}^{(0+1)}_{ij,V/A} 
\langle{\cal O}\rangle \right]^{(D=8)}
         \over M_\tau^8} \, , 
\eeqn
and
$\delta^{(2n)}_{ij,V/A} \simeq 0$   for $2 n \not= 6,8$.         
 
When the logarithmic dependence
of the Wilson coefficients on $s$ is taken into account, operators of
dimensions other than 6 and 8 do
contribute, but they are suppressed by two powers
of $\alpha_s(M_\tau^2)$.
The largest power corrections to $R_\tau$  then come from
dimension--6 operators, which have no such suppression.
Their size  was first estimated
in Ref.~\citen{BNP:92}, using published phenomenological fits to different
sets of data: 
\be\label{eq:delta_6}
\delta^{(6)}_{ij} \approx -{2\over 7}
\delta^{(6)}_{ij,V} \approx {2\over 11} 
\delta^{(6)}_{ij,A} \approx -(0.7\pm0.4)\% .
\ee
These power corrections are numerically
very small, which is due to the fact that they fall off
like the sixth power of $1/M_\tau$.
Moreover, there is a large cancellation 
between the vector and axial--vector contributions
to the total hadronic width
(the operator with the largest Wilson coefficient contributes with
opposite signs to the vector and axial--vector correlators, due to the
$\gamma_5$ flip). Thus, the
non-perturbative corrections to $R_\tau$ are smaller than the
corresponding contributions to  $R_{\tau,V/A}$.
A more detailed study of non-perturbative corrections, including
the very small $D=4$ contributions proportional to quark masses
or to $\alpha_s(M_\tau^2)^2$, can be found in Ref.~\citen{BNP:92}.

The estimate \eqn{eq:delta_6}
introduces a small uncertainty in the $R_\tau$ predictions,
since the actual evaluation of the non-perturbative contributions
involves a mixture of experimental
measurements and theoretical considerations, which are model--dependent
to some extent.
It is better to directly measure those contributions from the
$\tau$--decay data themselves. This information can be extracted 
\cite{SLAC:89} from
the invariant--mass distribution of the final hadrons in $\tau$ decay.

Although the distributions themselves cannot be predicted at present,
certain weighted integrals of the hadronic spectral functions can be
calculated in the same way as $R_\tau$.
The analyticity properties of $\Pi^{(J)}_{ij,V/A}$ imply \cite{SLAC:89,LDP:92b}:
%that, for any arbitrary weight function $W(s)$ without singularities in the
%region $|s|\leq s_0$, 
%
\bel{eq:weighted_integrals}
\int_0^{s_0} ds\, W(s)\: \mbox{\rm Im}\Pi^{(J)}_{ij,V/A}\, =\,
{i\over 2} \oint_{|s|=s_0} ds\, W(s) \,\Pi^{(J)}_{ij,V/A}\, , 
\ee
with $W(s)$ an arbitrary weight function without singularities in the
region $|s|\leq s_0$.
Generally speaking, the accuracy of the theoretical predictions can be
much worse than the one of $R_\tau$, because non-perturbative effects
are not necessarily suppressed.
In fact, choosing an appropriate weight function, non-perturbative effects
can even be made to dominate the final result. But this is precisely
what makes these integrals interesting: they can be used to measure the
parameters characterizing the non-perturbative dynamics.

To perform an experimental analysis, it is convenient to use
moments of the directly measured invariant--mass distribution
\cite{LDP:92b} ($k,l\ge 0$)
\be\label{eq:moments}
R^{kl}_\tau(s_0) \equiv\int_0^{s_0}\, ds
 \, \left(1 - {s\over s_0}\right)^k \left ( {s \over M_\tau^2} \right )^l
{ d R_\tau \over d s} \, .
\ee
The factor $(1-s/s_0)^k$ supplements $(1-s/M_\tau^2)^2$ for $s_0\not= M_\tau^2$,
in order to squeeze the integrand at the crossing of the positive real axis
and, therefore, improves the reliability of the OPE analysis; moreover, for
$s_0=M_\tau^2$ it reduces the contribution from the tail of the distribution,
which is badly defined experimentally.
A combined fit of
different $R_\tau^{kl}(s_0)$ moments
results in experimental
values for $\alpha_s(M_\tau^2)$ 
and for the coefficients of the inverse power corrections
in the OPE.
$R_\tau^{00}(M_\tau^2) = R_\tau$ 
uses the overall normalization of the hadronic
distribution, while the ratios 
$D_\tau^{kl}(M_\tau^2) = R^{kl}_\tau (M_\tau^2)/R_\tau$ are based on
the shape of the $s$ distribution and are more dependent on 
non-perturbative effects \cite{LDP:92b}.

The predicted suppression \cite{BR:88,NP:88,ORSAY:90,BNP:92}
of the non-perturbative corrections has been confirmed by
ALEPH \cite{ALEPH:93,DU:95} and CLEO \cite{CLEO:95},
using the moments (0,0), (1,0), (1,1), (1,2) and (1,3).
The most recent ALEPH analysis \cite{Hocker} gives:
\bel{eq:del_np}
\delta_{\mbox{\rms NP}} \equiv \sum_{D\geq 4} 
\left( \cos^2 \theta_C \delta^{(D)}_{ud}
         + \sin^2 \theta_C \delta^{(D)}_{us} \right)
         %\delta^{(D)} =
= -(0.5\pm 0.7)\% \, ,
\ee
in agreement with \eqn{eq:delta_6}.

\section{Phenomenology}

The QCD prediction
for $R_\tau$ is then completely dominated by the
perturbative contribution $\delta^{(0)}$; 
non-perturbative effects being of the order of
the perturbative uncertainties from uncalculated higher--order
corrections \cite{QCD:94,NA:95,BR:96}. Furthermore, 
as shown in Table~\ref{tab:perturbative}, the result turns out to be
very sensitive to the value of $\alpha_s(M_\tau^2)$, allowing for an accurate
determination of the fundamental QCD coupling.

The experimental value for $R_\tau$ can be obtained
from the leptonic branching fractions or from the
$\tau$ lifetime. The average of those determinations \cite{tau96}
\be  
  R_\tau =3.649 \pm 0.014 \, ,
\ee
corresponds to 
\be\label{eq:alpha}
\alpha_s(M_\tau^2)  =  0.355\pm 0.025 \, . 
\ee
%

%%%%%%%%%%%%%%%%%% FIGURE %%%%%%%%%%%%%%%%%%%%%
\begin{figure}[tbh]
\centerline{\epsfxsize =5.5cm \epsfbox{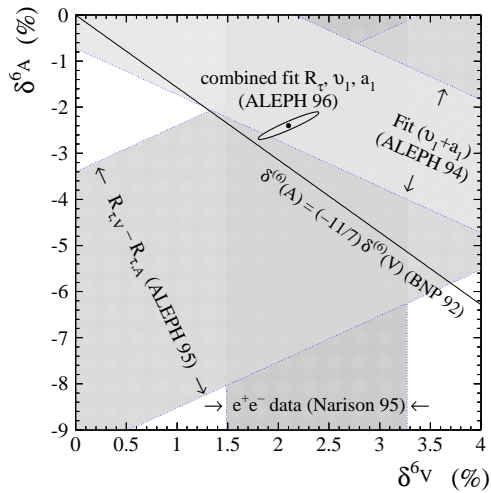}}
\vspace{-1.0cm}
\caption{Constraints on 
$\delta^{(6)}_V$ and $\delta^{(6)}_A$ obtained from
ALEPH data \protect\cite{Hocker}.
The ellipse depicts the combined fit.
All results are still preliminary. 
(Taken from Ref.~\protect\citen{Hocker})\hfill\mbox{}}
\label{fig:delta_6}
\end{figure}
%%%%%%%%%%%%%% END FIGURE %%%%%%%%%%%%%%%%%%%%%

Once the running coupling constant $\alpha_s(s)$ is determined at the scale
$M_\tau$, it can be evolved to higher energies using the renormalization
group.  The size of its error bar scales roughly
as $\alpha_s^2$, and it therefore shrinks as the scale increases.
Thus a modest precision in the
determination of $\alpha_s$ at low energies results in a very high
precision in the coupling constant at high energies.
After evolution up to the scale $M_Z$, the strong coupling constant in
\eqn{eq:alpha} decreases to \footnote{
%%%%%%%%%%%
From a combined analysis of $\tau$ data, ALEPH quotes
\protect\cite{Hocker}:
$\alpha_s(M_Z^2)  =
0.1225\pm 0.0006_{\mbox{\protect\rms exp}}\pm
0.0015_{\mbox{\protect\rms th}}\pm 0.0010_{\mbox{\protect\rms evol}}$.  
%0.1229\pm 0.0011_{\mbox{\protect\rms exp}}\pm
%0.0014_{\mbox{\protect\rms th}}\pm 0.0010_{\mbox{\protect\rms evol}}$.
}
%%%%%%%%%%%% 
%
\be\label{eq:alpha_z}
\alpha_s(M_Z^2)  =  0.1225\pm 0.0025 \, ,
\ee
in excellent
agreement with the direct measurement \cite{SC:97,BE:96} at $\mu=M_Z$,
%present LEP average \cite{SC:97,BE:96} (without $R_\tau$)
$\alpha_s(M_Z^2)  =  0.121\pm0.003$,  
and with a similar error bar.
The comparison of these two determinations of $\alpha_s$ in two extreme
energy regimes, $M_\tau$ and $M_Z$, provides a beautiful test of the
predicted running of the QCD coupling.

With $\alpha_s(M_\tau^2)$ fixed to the value in Eq.~(\ref{eq:alpha}), 
the same theoretical framework gives definite
predictions \cite{BNP:92,QCD:94} for the semi-inclusive $\tau$ decay widths
$R_{\tau,V}$, $R_{\tau,A}$ and $R_{\tau,S}$, in good agreement with the
experimental measurements \cite{Hocker,Davier}.
The separate analysis of the vector and axial--vector contributions
allows to investigate the associated non-perturbative corrections
($R_{\tau,V}-R_{\tau,A}$ is a pure non-perturbative quantity).
Figure~\ref{fig:delta_6} shows \cite{Hocker}
the (preliminary) constraints on 
$\delta^{(6)}_V$ and $\delta^{(6)}_A$ obtained from the most recent
ALEPH analyses \cite{ALEPH:93,Hocker,ALEPH:96}.
A clear improvement over previous phenomenological determinations 
\cite{BNP:92,NA:95b} is apparent.

%%%%%%%%%%%%%%%%%% FIGURE %%%%%%%%%%%%%%%%%%%%%
\begin{figure}[tbh]
\centerline{\epsfxsize =7.3cm \epsfbox{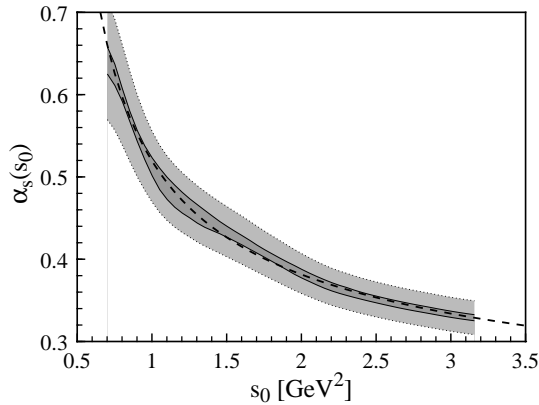}}
%\vspace{-1.0cm}
\caption{Values of $\alpha_s(s_0)$ extracted 
\protect\cite{GN:96} from the $R_\tau^{00}(s_0)$ data 
\protect\cite{DU:95,CLEO:95}. 
The dashed line shows the three--loop QCD
prediction for the running coupling constant.
(Taken from Ref.~\protect\citen{GN:96})\hfill\mbox{}}
\label{fig:alphang}
\end{figure}
%%%%%%%%%%%%%% END FIGURE %%%%%%%%%%%%%%%%%%%%%

The Cabibbo--suppressed width $R_{\tau,S}$ is very
sensitive to the value of the strange quark mass \cite{BNP:92},
providing a direct and clean way of measuring $m_s$.
A very preliminary value,
$m_s(M_\tau^2)= (212\, {}^{+30\, +1}_{-35\, -5})$ MeV,
has been already presented at the
last tau workshop \cite{Davier}.

Using the measured invariant--mass distribution of the final hadrons, it
is possible to evaluate the integral $R_\tau^{00}(s_0)$,
with an arbitrary
upper limit of integration $s_0\leq M_\tau^2$. The experimental $s_0$
dependence  agrees well with the
theoretical predictions \cite{LDP:92b} up to rather low values\cite{GN:96}
of $s_0$ ($> 0.7$ GeV$^2$). 
Equivalently,  from the measured \cite{DU:95,CLEO:95}
$R_\tau^{00}(s_0)$ distribution one obtains
$\alpha_s(s_0)$ as a function of the scale $s_0$.
As shown  \cite{GN:96} in Figure~\ref{fig:alphang},
the result exhibits an impressive agreement with the
running predicted at three--loop order by QCD.
It is important to realize \cite{GN:96}
that the theoretical prediction for
$R_\tau^{00}(s_0)$ does not contain inverse powers of $s_0$
(as long as the $s$--dependence of the Wilson coefficients is ignored).
The power corrections are suppressed by powers of $1/M_\tau^2$;
thus, they do not drive a break--down of the OPE. This could explain
the surprisingly good agreement with the data for $s_0\lsim 1$ GeV$^2$

A similar test was performed before \cite{NP:93} for $R_{\tau,V}$, 
using the vector spectral
function measured in $e^+e^-\to$ hadrons, and
varying the value of the tau mass. This allows to
study the behaviour of the OPE at lower scales.
The theoretical predictions for $R_{\tau,V}$ as function of $M_\tau^2$
agree \cite{NP:93} well with the data for $M_\tau > 1.2$ GeV.
Below this value, higher--order inverse power corrections become
very important and eventually generate the expected break--down
of the expansion in powers of $1/M_\tau^2$.

\section{Summary}

Because of its inclusive nature,
the total hadronic width of the $\tau$ can be rigorously computed
within QCD.
One only needs to study two--point 
correlation functions for the vector and axial--vector currents.
As shown in Eq.~(\ref{eq:circle}),
this information is only needed in the complex plane,
away from the time--like axis; the dangerous region near the physical
cut does not contribute at all to the result, because of the phase--space
factor $(1-s/M_\tau^2)^2$.
The uncertainties of the theoretical predictions are then quite small.

The ratio $R_\tau$ is very
sensitive to the value of the strong coupling, and therefore can be
used \cite{NP:88} to measure $\alpha_s(M_\tau^2)$.
This observation has triggered an ongoing effort to improve
the knowledge of $R_\tau$ from both  the experimental and the theoretical
sides.
The fact that $M_\tau$ is a quite low energy scale 
(i.e. that $\alpha_s(M_\tau^2)$ is big),
but still large enough to allow a perturbative analysis, makes 
$R_\tau$ an ideal observable to determine the QCD coupling.
Moreover, since the error of $\alpha_s(\mu^2)$ shrinks as $\mu$ increases,
the good accuracy of the $R_\tau$ determination of $\alpha_s(M_\tau^2)$ implies
a very precise value  of $\alpha_s(M_Z^2)$.

The theoretical analysis of $R_\tau$ has reached a very mature level.
Many different sources of possible perturbative and non-perturbative
contributions have been analyzed in detail. 
A very detailed study of the associated uncertainties has been given
in Ref.~\citen{QCD:94}.
The final theoretical
uncertainty is small and has been adequately taken into account in the
final $\alpha_s(M_\tau^2)$ determination in Eq.~(\ref{eq:alpha}). 

The comparison of the theoretical  predictions with the experimental
data shows
a successful and consistent picture.
The $\alpha_s(M_\tau^2)$ determination is in excellent agreement with
the measurements at the $Z$--mass scale, 
providing clear evidence of the running of
$\alpha_s$. Moreover, the analysis of the semi-inclusive components
of the  $\tau$ hadronic width, $R_{\tau,V}$, $R_{\tau,A}$ 
and $R_{\tau,S}$, and the invariant--mass distribution of the final
decay products
gives a further confirmation of the reliability 
of the theoretical
framework, and allows to investigate other important
QCD parameters such as the strange--quark mass or the non-perturbative
vacuum condensates.

\section*{Acknowledgments}
I would like to thank the organizers for this enjoyable meeting,
and specially Matthias Jamin for his kind hospitality.
I am also indebted to Andreas H\"ocker and Matthias Neubert
for providing
%the original postscript file with 
figures~\ref{fig:delta_6} and \ref{fig:alphang}.
This work has been supported in part by CICYT (Spain) under grant 
No. AEN-96-1718.

\end{document}